\documentclass[prd,notitlepage,longbibliography,nofootinbib,superscriptaddress,onecolumn,preprintnumbers]{revtex4-2}
\usepackage[utf8]{inputenc}

\usepackage{bm}
\usepackage{comment} 
\usepackage[colorlinks=true,urlcolor=blue,anchorcolor=black,citecolor=blue,linkcolor=red,filecolor=black,menucolor=black,pagecolor=black,linktocpage=true,pdfproducer=medialab,pdfa=true]{hyperref}
\usepackage{graphicx}
\usepackage{amsmath,latexsym,amssymb,mathrsfs,ascmac,mathtools}
\usepackage{multirow}
\usepackage{braket}
\usepackage{placeins}
\begin{document}

\title{Quantum Dynamics of the Schwarzschild Interior in Ashtekar–Barbero Variables with Minimal Length Effects}

\author{Takamasa Kanai}
\email{kanai@kochi-ct.ac.jp}

\affiliation{Department of Social Design Engineering,
National Institute of Technology (KOSEN), Kochi College,
200-1 Monobe Otsu, Nankoku, Kochi, 783-8508, Japan}

\begin{abstract}
We study the quantum dynamics of the Schwarzschild interior in the Ashtekar–Barbero formulation, focusing on the fate of the classical singularity and the annihilation-to-nothing scenario. Using minisuperspace Wheeler–DeWitt quantization, we first analyze the standard Schrödinger representation and show that the annihilation-to-nothing behavior appears only for a specific choice of factor ordering and is not generic.

We then introduce a generalized uncertainty principle (GUP), which induces minimal-length effects through a deformation of the canonical algebra. Solving the modified Wheeler–DeWitt equation and constructing Gaussian wave packets localized at the horizon, we find that the annihilation-to-nothing behavior is suppressed once the GUP corrections are included. Our results indicate that minimal-length effects qualitatively alter the quantum interior dynamics and challenge the robustness of this scenario as a mechanism for singularity resolution.
\end{abstract}
\maketitle

\section{Introduction}

The fate of spacetime singularities remains one of the central open problems in quantum gravity. In classical general relativity, the Schwarzschild solution inevitably develops a curvature singularity inside the event horizon \cite{Penrose:1964wq,Hawking:1970zqf,Hawking:1973uf}, where invariants such as the Kretschmann scalar diverge and the classical description breaks down. A consistent quantum theory of gravity is widely expected to modify this behavior and provide a mechanism for singularity resolution.

A useful arena for addressing this issue is the interior region of the Schwarzschild black hole. Owing to its isometry with a Kantowski–Sachs cosmological spacetime, the system admits a minisuperspace description with finitely many degrees of freedom. Within this framework, the Wheeler–DeWitt (WDW) equation provides a canonical approach to quantum dynamics \cite{Halliwell:1989myn,Kiefer:2013jqa,Kiefer:2008sw,Kiefer:2025udf}. Various studies have investigated whether appropriate boundary conditions or wave-packet constructions can eliminate the classical singularity at the quantum level.

In particular, the annihilation-to-nothing scenario proposed suggests a novel interpretation of the WDW wave function inside the horizon \cite{Bouhmadi-Lopez:2019kkt,Yeom:2019csm,Yeom:2021bpd,Brahma:2021xjy}. In this picture, wave packets corresponding to classical branches with opposite time orientations mutually annihilate before reaching the singularity, leading to the disappearance of the interior spacetime. This mechanism has been discussed in the conventional metric formulation of the minisuperspace model and has been suggested as a possible quantum resolution of the Schwarzschild singularity.

However, it should be emphasized that general relativity is widely regarded as a low-energy effective theory of a more fundamental quantum theory of gravity \cite{Weinberg:1978kz,Donoghue:1993eb,Donoghue:1994dn,Burgess:2003jk}. As such, it does not incorporate the ultraviolet (UV) dynamics or additional physical degrees of freedom that are expected to become relevant near Planckian scales. From this perspective, it would be surprising if a complete resolution of spacetime singularities could be achieved entirely within a framework that lacks genuine UV structure. In particular, mechanisms for singularity resolution are generally expected to arise from new high-energy dynamics or novel microscopic degrees of freedom absent in classical general relativity.

Indeed, it has been argued in Ref.~\cite{Kanai:2025mrx} that the annihilation-to-nothing scenario breaks down when analyzed strictly within a low-energy effective framework, and that its validity can only be properly assessed in a UV-complete theory. This observation suggests that conclusions drawn from the undeformed WDW equation, treated as an effective description, may not be sufficient to establish the robustness of the scenario. Therefore, whether the annihilation-to-nothing proposal provides a consistent mechanism for singularity resolution should be examined in a setting that incorporates ultraviolet dynamics or additional high-energy degrees of freedom beyond those contained in general relativity.

The purpose of the present work is twofold. First, we reformulate the Schwarzschild interior dynamics in terms of Ashtekar–Barbero connection–triad variables. This formulation provides a canonical structure closely related to loop quantum gravity (LQG) \cite{Rovelli:1994ge,Rovelli:1997yv,Ashtekar:2017yom}. Applications of LQG techniques to black hole interiors have been widely investigated in the literature \cite{Modesto:2004wm,Ashtekar:2005qt,Corichi:2015xia,Sartini:2020ycs,Geiller:2020xze,Ongole:2022rqi}, and this framework allows us to clarify the role of factor ordering in the WDW equation. We show that the realization of the annihilation-to-nothing behavior depends sensitively on the choice of ordering parameter and is not generic within the Ashtekar–Barbero framework.

Second, we investigate the impact of ultraviolet-motivated corrections on this scenario by incorporating a generalized uncertainty principle (GUP) \cite{Maggiore:1993rv,Kempf:1994su,Scardigli:2003kr,Mignemi:2011wh,Hossenfelder:2012jw,Pramanik:2013zy,Bosso:2020aqm,Wagner:2021bqz,Barca:2021epy,Bosso:2021koi,HerkenhoffGomes:2022bnh,Segreto:2022clx,Bosso:2022rue}. The GUP introduces minimal uncertainties in the conjugate momenta through a deformation of the canonical algebra, effectively encoding minimal-length effects expected from quantum gravity such as string theory \cite{Polchinski:1998rq,Polchinski:1998rr} and other approaches to quantum gravity. Implementing this deformation in the reduced phase space, we derive the modified WDW equation and obtain analytic expressions for its solutions. By constructing Gaussian wave packets localized at the horizon, we analyze whether the annihilation-to-nothing behavior persists in the presence of minimal-length effects.

Our results indicate that while the conventional quantization can reproduce the annihilation-to-nothing behavior for a specific factor ordering, the inclusion of GUP corrections suppresses this structure across the range of orderings considered. This suggests that the scenario is not robust under ultraviolet modifications and that minimal-length effects qualitatively alter the quantum dynamics of the Schwarzschild interior.

This paper is organized as follows. In Sec.~\ref{review Yoem}, we review the minisuperspace quantization of the Schwarzschild interior and the original formulation of the annihilation-to-nothing scenario. In Sec.~\ref{annihilation}, we present the Ashtekar–Barbero formulation of the Schwarzschild interior and analyze the resulting WDW quantization in the absence of GUP corrections. In Sec.~\ref{GUP}, we incorporate the generalized uncertainty principle and derive the corresponding modified quantum dynamics. Sec.~\ref{summary} is devoted to conclusions and discussion.

Unless specified otherwise, we use the natural unit system, where the speed of light $c=1$ and the Dirac constant $\hbar=1$.

\section{Minisuperspace Quantization of Black Hole Interiors and Yeom's Proposal}
\label{review Yoem}

In this section, we briefly summarize the minisuperspace quantization of the interior region of a spherically symmetric black hole and review Yeom's interpretation of the resulting Wheeler–DeWitt (WDW) wave function.

\subsection{Minisuperspace formulation inside the horizon}

For vacuum, spherically symmetric configurations, Birkhoff’s theorem \cite{Misner:1973prb} ensures that the geometry is uniquely described by the Schwarzschild–Tangherlini solution in $D$ spacetime dimensions. Consequently, the interior of such a black hole provides a universal setting for analyzing minisuperspace quantum dynamics.

We begin with the $D$-dimensional Einstein–Hilbert action,
\begin{equation}
S = \frac{1}{16\pi G}\int d^Dx \sqrt{-g} R .
\end{equation}

Inside the horizon, the roles of temporal and radial coordinates are interchanged. A convenient parametrization is given by the Kantowski–Sachs-type metric \cite{Kantowski:1966te},
\begin{equation}
ds^2 = - e^{2n(t)} dt^2 + e^{2\alpha(t)} dr^2
+ r_s^2 e^{2\beta(t)} d\Omega_{D-2}^2 ,
\end{equation}
where $r_s$ denotes a constant length scale and $d\Omega_{D-2}^2$ is the unit $(D-2)$-sphere metric.

Matching to the Schwarzschild–Tangherlini interior yields
\begin{equation}
e^{2\beta} = \frac{t^2}{r_s^2}, \qquad
e^{2\alpha} = \frac{r_s^{D-3}}{t^{D-3}} - 1, \qquad
e^{2n} = \left(\frac{r_s^{D-3}}{t^{D-3}} - 1\right)^{-1}.
\end{equation}

Choosing the lapse gauge
\begin{equation}
n(t)=\alpha(t)+(D-2)\beta(t),
\end{equation}
the reduced action becomes
\begin{equation}
S \propto \int dt
\left[
-2\dot{\alpha}\dot{\beta}
-(D-3)\dot{\beta}^2
+(D-3) r_s^{2(D-3)} e^{2\alpha+2(D-3)\beta}
\right].
\end{equation}

Introducing new variables
\begin{equation}
X=\alpha, \qquad
Y=\alpha+(D-3)\beta,
\end{equation}
diagonalizes the kinetic term, leading to
\begin{equation}
S \propto \int dt
\left[
\frac{1}{D-3}(\dot{X}^2-\dot{Y}^2)
+(D-3) r_s^{2(D-3)} e^{2Y}
\right].
\end{equation}

The corresponding Hamiltonian reads
\begin{equation}
H=(D-3)\left[
\frac14(\Pi_X^2-\Pi_Y^2)
-r_s^{2(D-3)}e^{2Y}
\right].
\end{equation}

Upon canonical quantization,
\begin{equation}
\Pi_X \rightarrow -i\partial_X,
\qquad
\Pi_Y \rightarrow -i\partial_Y,
\end{equation}
and imposing the Hamiltonian constraint $\hat{\mathcal H}\Psi=0$, we obtain the minisuperspace WDW equation \cite{Halliwell:1989myn,Kiefer:2013jqa,Kiefer:2008sw,Kiefer:2025udf},
\begin{equation}
\left(
\frac{\partial^2}{\partial X^2}
-
\frac{\partial^2}{\partial Y^2}
+
4 r_s^{2(D-3)} e^{2Y}
\right)
\Psi(X,Y)=0.
\end{equation}

The structure of this equation is universal, with dimensional dependence entering only through $r_s$.

In this parametrization, the limit $X,Y\to -\infty$ corresponds to the horizon, whereas $X\to\infty$ with $Y\to-\infty$ approaches the classical singularity.

\subsection{Wave-packet solutions and annihilation-to-nothing}

A separable solution of the WDW equation is given by
\begin{equation}
\label{sol1}
\psi_k(X,Y)
=
e^{-ikX}K_{ik}(2r_s e^{Y}),
\end{equation}
where $K_{ik}$ denotes the modified Bessel function of the second kind.

General states are superpositions
\begin{equation}
\Psi(X,Y)
=
\int_{-\infty}^{\infty}
f(k)\psi_k(X,Y)\,dk.
\end{equation}

\begin{figure}[htbp]
   \centering
   \includegraphics[width=0.7\linewidth]{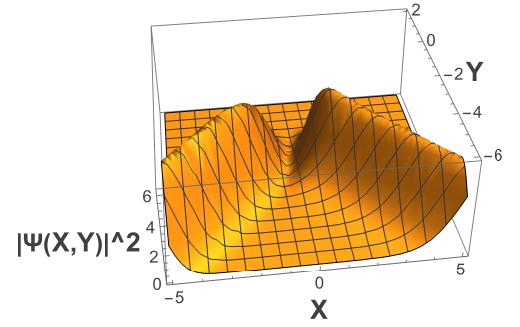}
   \caption{Probability density of the wave function for $C=\sigma=r_s=1$. The integral over the wave number $k$ is taken over the interval $[-8,8]$. Adapted from Ref.~\cite{Bouhmadi-Lopez:2019kkt}.}
   \label{fig1}
\end{figure}

Appropriate choices of the weight function $f(k)$ allow one to construct localized wave packets. For instance, choosting $f(k)$ as a Gaussian wave packet such that
\begin{align}
\label{amplitude1}
f(k)=\frac{2Ce^{-\sigma^2k^2/2}}{\Gamma(-ik)r_s^{ik}},
\end{align}
where $C$ is the normalization constant and $\sigma$ is the standard deviation of the pulse at $X=Y$.Then, the wave packet is
\begin{equation}
\Psi_1
=\int_{-\infty}^{\infty}\frac{2Ce^{-\sigma^2k^2/2}}{\Gamma(-ik)r_s^{ik}}e^{-ikX}K_{ik}\left(2r_se^Y\right)dk.
\end{equation}
To investigate the formation and possible resolution of the interior singularity, we superpose a Gaussian wave packet exhibiting classical behavior at the horizon. The corresponding probability density peaks along the classical trajectory while it vanishes near $X=0$. Following \cite{Bouhmadi-Lopez:2019kkt,Yeom:2019csm,Yeom:2021bpd,Brahma:2021xjy}, this feature has been interpreted as an ``annihilation-to-nothing'' phenomenon, describing the quantum annihilation of two interior classical branches with opposite time orientations. In this scenario, wave packets propagating along opposite arrows of time undergo mutual annihilation inside the horizon, rather than in the vicinity of the singularity. As a result, the black hole spacetime disappears, thereby resolving the classical singularity. A graphical representation of this behavior is shown in Fig.~\ref{fig1}.

\section*{Overview}

This paper studies the quantization of the Schwarzschild black hole interior by comparing two different approaches:
\begin{itemize}
    \item Standard canonical quantization
    \item Quantization based on a minimal uncertainty principle (GUP-inspired approach)
\end{itemize}

The interior of the Schwarzschild black hole is isometric to a Kantowski--Sachs cosmological spacetime, which allows the system to be treated
as a minisuperspace model with finitely many degrees of freedom. The classical theory is formulated in terms of Ashtekar--Barbero variables.

The main objective of this study is to examine whether the classical singularity is resolved at the quantum level. In particular, we investigate the “annihilation-to-nothing” scenario as one possible hypothesis for singularity resolution.

\section{Ashtekar–Barbero Formulation of the Annihilation-to-Nothing Scenario}
\label{annihilation}

In this section, we reformulate the annihilation-to-nothing scenario proposed by Yeom et al. within the framework of the Ashtekar-Barbero variables. Our goal is to reconstruct the quantum dynamics of the Kantowski–Sachs–type geometry in terms of connection-triad variables, thereby providing a canonical description suitable for the Wheeler-DeWitt quantization. This reformulation allows us to clarify the role of the canonical structure and to examine the scenario from the viewpoint of loop-inspired variables.

As a first step, we quantize the Kantowski-Sachs-type model using the standard Schr\"odinger representation in order to compare the result
with the GUP-based quantization presented later.

\subsection{Ashtekar-Barbero formulation of the Kantowski-Sachs geometry}

In this section, we investigate the annihilation-to-nothing scenario in the Ashtekar-Barbero framework. To make the analysis self-contained, we first review the essential ingredients of the Ashtekar-Barbero formulation of the Kantowski-Sachs type model.

As is well-known, the metric of the interior of the Schwarzschild black hole can be obtained by exchanging $t\leftrightarrow r$ in the Schwarzschild coordinates, yielding
\begin{align}
ds^2=-\left(\frac{2GM}{t}-1\right)^{-1}dt^2+\left(\frac{2GM}{t}-1\right)dr^2+t^2(d\theta^2+\sin^2\theta d\phi^2).
\end{align}

This line element describes a homogeneous but anisotropic spacetime of Kantowski--Sachs type, where the coordinate $t$ now plays the role of an areal radius and becomes a timelike variable inside the horizon. The spatial topology of constant-$t$ slices is $\mathbb{R}\times S^2$, reflecting the radial direction and the two-sphere symmetry.

To quantize this system within the Ashtekar--Barbero formalism, we exploit the symmetry reduction associated with the Kantowski--Sachs geometry. The invariance under spatial diffeomorphisms generated by a vector field $\xi^a$ implies that the connection $A^i_a$ and densitized triad $E^a_i$ transform according to
\begin{align}
\pounds_\xi A=D\Lambda,\qquad \pounds_\xi E=[E,\Lambda],
\end{align}
where $D$ denotes the gauge-covariant derivative and $\Lambda$ is an $\mathfrak{su}(2)$-valued gauge parameter. These conditions implement symmetry reduction up to internal gauge transformations.

Imposing these symmetry conditions and adapting the variables to the fiducial cell of length $L_0$ in the non-compact radial direction, the most general invariant connection and triad compatible with the Kantowski--Sachs symmetry can be parametrized by two canonical pairs $(b,p_b)$ and $(c,p_c)$. Explicitly, they take the form
\begin{align}
A^i_a\tau_i dx^a&=\frac{c}{L_0}\tau_3 dr
+b\,\tau_2 d\theta
-b\,\tau_1\sin\theta d\phi
+\tau_3\cos\theta d\phi,\\
E_i^a\tau_i\partial_a&=
p_c\,\tau_3\sin\theta\,\partial_r
+\frac{p_b}{L_0}\tau_2\sin\theta\,\partial_\theta
-\frac{p_b}{L_0}\tau_1\,\partial_\phi,
\end{align}
where $\tau_i=-\frac{i}{2}\sigma_i$ are the generators of $\mathfrak{su}(2)$ written in terms of the Pauli matrices \cite{Ashtekar:2005qt}. The variables $b$ and $c$ encode the connection components along the angular and radial directions, respectively, while $p_b$ and $p_c$ determine the corresponding triad components and are directly related to the physical metric degrees of freedom.

Using Ashtekar--Barbero variables $(b,c)$ and their conjugate momenta
$(p_b,p_c)$, the Hamiltonian constraint of the Schwarzschild interior
is obtained. After inserting the symmetry-reduced connection and triad
into the Einstein-Hilbert action and performing the Legendre transform,
one arrives at the reduced Hamiltonian
\begin{align}
H=-\frac{N\mathrm{sgn}(p_c)}{2G\gamma^2}\left[2bc\sqrt{|p_c|}+(b^2+\gamma^2)\frac{p_b}{\sqrt{|p_c|}}\right],
\end{align}
where $\gamma$ denotes the Barbero--Immirzi parameter and $N$ is the lapse function \cite{Ashtekar:2005qt}.
The dynamics is generated through the canonical Poisson brackets
\begin{align}
\{ c,p_c\}=2G\gamma,\qquad 
\{ b,p_b\}=G\gamma,
\end{align}
so that $(b,p_b)$ and $(c,p_c)$ form two independent canonical pairs.

The Kantowski-Sachs-type metric in these variables is given by
\begin{align}
ds^2=-N(T)^2dT^2+\frac{p_b^2}{L_0^2|p_c|}dr^2+|p_c|(d\theta^2+\sin^2\theta d\phi^2),
\end{align}
where $p_c$ determines the physical area of the two-spheres, $A_{S^2}=4\pi |p_c|$, while $p_b$ controls the radial component of the metric. In particular, the classical Schwarzschild solution is recovered by solving Hamilton's equations derived from $H\approx0$ and choosing an appropriate lapse function.

Under the rescaling
\begin{align}
p_b \to L_0\, p_b, \qquad c \to L_0\, c,
\end{align}
the canonical commutators are merely rescaled by constant factors, and the resulting Wheeler--DeWitt equation retains its original form. Therefore, one can, without loss of generality, set \(L_0 = 1\). In what follows, we adopt this choice.

From the metric it follows that the classical singularity corresponds to the limit in which the two-sphere shrinks to zero area,
\begin{align}
p_b\rightarrow0,\qquad p_c\rightarrow0.
\end{align}
In contrast, the classical event horizon is characterized by a vanishing radial metric component while the two-sphere retains a finite area, which corresponds to
\begin{align}
p_b \rightarrow 0,\qquad p_c \rightarrow r_s^2,
\end{align}
where $r_s=2GM$ denotes the Schwarzschild radius. Thus, while both the horizon and the singularity involve $p_b\to 0$, they are distinguished by the behavior of $p_c$: it remains finite at the horizon but collapses to zero at the curvature singularity.

This geometric distinction is also reflected in the behavior of curvature invariants. 
In the singular limit the spacetime curvature grows without bound, signaling the breakdown of the classical description.

In this regime curvature invariants diverge. In particular, the Kretschmann scalar, which measures the squared magnitude of the Riemann
tensor, takes the form \cite{Rastgoo:2022mks}
\begin{equation}
K = \frac{12}{\gamma^4}
\frac{(b^2+\gamma^2)^2}{p_c^2},
\end{equation}
and therefore diverges as $p_c \to 0$. This divergence signals the presence of the classical spacetime singularity in the Schwarzschild
interior.

To analyze the quantum dynamics of the Schwarzschild interior and to examine the scenario proposed by Yeom et al., we now formulate the system in Hamiltonian form and proceed to its canonical quantization.

\subsection{Choice of Lapse Function}

In the classical theory, the choice of lapse function $N$ does not alter the physical content. A convenient choice that simplifies the equations of motion by decoupling $(c,p_c)$ from $(b,p_b)$ is
\begin{equation}
N = \mathrm{sgn}(p_c)\, |p_c|.
\end{equation}

With this choice, the Hamiltonian constraint reduces to
\begin{equation}
H = -\frac{1}{2G\gamma^2}
\left[
(b^2+\gamma^2)p_b + 2cbp_c
\right].
\label{eq:Hamiltonian_standard}
\end{equation}

Having obtained the classical Hamiltonian constraint in terms of the canonical variables, we are now in a position to apply the Dirac quantization procedure.

\subsection{Canonical Quantization}

We impose the Dirac quantization prescription on the classical Poisson algebra:
\begin{align}
[b,p_b] &= iG\gamma, \\
[c,p_c] &= 2iG\gamma, \\
[b,p_c] &= [c,p_b] = 0.
\end{align}

In the Schr\"odinger representation, where $b$ and $c$ act multiplicatively,
the momenta are represented as differential operators:
\begin{align}
p_b &= -iG\gamma \frac{\partial}{\partial b}, \\
p_c &= -2iG\gamma \frac{\partial}{\partial c}.
\end{align}

Substituting these operator representations into the Hamiltonian constraint yields the Wheeler-DeWitt equation for the wave function $\psi(b,c)$.

\subsection{Hamiltonian Constraint Equation}

Acting on a generic wave function $\psi(b,c)$,
the Hamiltonian constraint becomes a differential equation.
Allowing for a general factor ordering parametrized by
a dimensionless parameter $a$, the equation reads
\begin{equation}
(b^2+\gamma^2)^{1-a}
\frac{\partial}{\partial b}
\left[
(b^2+\gamma^2)^a \psi
\right]
+
4b\, c^{\,1-a}
\frac{\partial}{\partial c}
\left[
c^a \psi
\right]
= 0.
\label{eq:WDW_standard}
\end{equation}

Special choices of the ordering parameter are:
\begin{itemize}
\item $a = \tfrac{1}{2}$ : symmetric ordering,
\item $a = 1$ : configuration variables to the right of momenta,
\item $a = 0$ : momenta to the right of configuration variables.
\end{itemize}

\subsection{Separation of Variables}

Assuming a separable ansatz
\begin{equation}
\psi(b,c) = S(b)\,R(c),
\end{equation}
the Hamiltonian constraint splits into two independent ordinary
differential equations, one depending only on $b$ and the other only on
$c$. More explicitly, the action of the Hamiltonian constraint operator
on a generic state leads to the partial differential equation
\begin{equation}
(b^2+\gamma^2)^{1-a}\frac{\partial}{\partial b}
\left[(b^2+\gamma^2)^a \psi(b,c)\right]
+
4b\, c^{\,1-a}\frac{\partial}{\partial c}
\left[c^a \psi(b,c)\right]
=0,
\end{equation}
where $a$ parametrizes the chosen factor ordering \cite{Hartle:1983ai}.

Substituting the separable ansatz $\psi(b,c)=S(b)R(c)$ into the above
equation and dividing by $S(b)R(c)$, the dependence on $b$ and $c$
separates into two terms, each of which must be equal to a constant.
Introducing a dimensionless separation constant $m$, one obtains the
pair of ordinary differential equations
\begin{equation}
(b^2+\gamma^2)\frac{dS(b)}{db}
+
b(2a-m)S(b)
=0,
\end{equation}
\begin{equation}
4c\frac{dR(c)}{dc}
+
(4a+m)R(c)
=0.
\end{equation}

These equations can be integrated straightforwardly. The first equation
is a first-order linear differential equation whose solution is given by
\begin{equation}
S_m(b)
=
C_{1}(m,a)
\,(b^2+\gamma^2)^{\frac{m}{2}-a},
\end{equation}
where $C_{1}(m,a)$ is an integration constant that may depend on the
ordering parameter $a$ and the separation constant $m$.

Similarly, the second equation yields
\begin{equation}
R_m(c)
=
C_{2}(m,a)
\,c^{-\frac{m}{4}-a},
\end{equation}
with $C_{2}(m,a)$ another integration constant. The full solution is therefore labeled by the continuous parameter $m$, which plays the role
of a quantum number characterizing the separated modes of the deformed Wheeler-DeWitt equation.

Combining the solutions obtained in the $b$- and $c$-sectors, the full
separable wave function takes the form
\begin{equation}
\label{wave function1}
\psi_m(b,c)
=
A(m,a)
(b^2+\gamma^2)^{\frac{m}{2}-a}
c^{-\frac{m}{4}-a},
\end{equation}
where the overall normalization constant is defined as
$A(m,a)=C_{1}(m,a)C_{2}(m,a)$.

To analyze the physical content of these states, it is convenient to
pass to the momentum representation. This is achieved by performing the
Fourier transform
\begin{equation}
\label{wave function2}
\bar{\psi}_m(p_b,p_c)
=
\int_{-\infty}^{\infty}
\int_{-\infty}^{\infty}
e^{-ip_b b}e^{-ip_c c}
\psi_m(b,c)\, db\, dc.
\end{equation}
Substituting~(\ref{wave function1}) into~(\ref{wave function2}) and
evaluating the integrals using standard integral representations of the
Gamma and modified Bessel functions, one finds
\begin{align}
\bar{\psi}_m(p_b,p_c)
&=
A(m,a)
\frac{\sqrt{\pi}\,2^{\frac{1}{2}(3-2a+m)}
\gamma^{\frac{1}{2}(1-2a+m)}
\Gamma\!\left(1-a-\frac{m}{4}\right)
\sin\!\left(\pi(a+\frac{m}{4})\right)}
{e^{i\frac{\pi}{8}(m+4a)}
\Gamma\!\left(a-\frac{m}{2}\right)}
\nonumber\\
&\quad
\times
[\operatorname{sgn}(p_c)+1]
|p_c|^{-1+a+\frac{m}{4}}
|p_b|^{-\frac{1}{2}(1-2a+m)}
K_{-\frac{1}{2}(1-2a+m)}\!\left(\gamma |p_b|\right),
\end{align}
where $K_{\nu}(x)$ denotes the modified Bessel function of the second kind. Since the wave function vanishes identically for $p_c<0$, we may, without loss of generality, restrict our attention to the sector $p_c>0$.

To facilitate comparison with the classical geometry of the Schwarzschild interior, we introduce the variables
\begin{align}
p_c=r_s^2 e^{2Y-2X},
\qquad
p_b=L_0 r_s e^Y,
\end{align}
which are adapted to the horizon and singularity structure of the spacetime. In the present work, we set $L_0=1$ without loss of generality. 
Expressed in terms of $(X,Y)$, the wave function becomes
\begin{align}
\psi_m(X,Y)
&=
A(m,a)
\frac{\sqrt{\pi}\,2^{\frac{1}{2}(5-2a+m)}
\gamma^{\frac{1}{2}(1-2a+m)}
\Gamma\!\left(1-a-\frac{m}{4}\right)
\sin\!\left(\pi(a+\frac{m}{4})\right)}
{r_s^{\frac{5}{2}-3a}
e^{i\frac{\pi}{8}(m+4a)}
\Gamma\!\left(a-\frac{m}{2}\right)}
\nonumber\\
&\quad
\times
e^{2(1-a-\frac{m}{4})X}
e^{(-\frac{5}{2}+3a)Y}
K_{-\frac{1}{2}(1-2a+m)}\!\left(r_s\gamma e^Y\right).
\end{align}

In order for the $X$-dependent part of the solution to take the plane-wave form consistent with~(\ref{sol1}), we introduce a new parameter $k$ through
\begin{align}
2\left(1-a-\frac{m}{4}\right) = -ik ,
\end{align}
so that the exponential dependence on $X$ becomes $e^{-ikX}$. Reexpressing the solution in terms of $k$, we obtain
\begin{align}
\psi_k(X,Y)
&=
\tilde{A}(k,a)
\frac{\sqrt{\pi}\,2^{\frac{1}{2}(9-6a+2ik)}
\gamma^{\frac{1}{2}(5-6a+ik)}
\Gamma\!\left(-\frac{ik}{2}\right)
\sin\!\left(\pi\left(1+\frac{ik}{2}\right)\right)}
{r_s^{\frac{5}{2}-3a}
e^{i\frac{\pi}{2}\left(1+\frac{ik}{2}\right)}
\Gamma\!\left(-2+3a-ik\right)}
\nonumber\\
&\quad
\times
e^{-ikX}
e^{(-\frac{5}{2}+3a)Y}
K_{-\frac{5}{2}+3a-ik}\!\left(r_s\gamma e^Y\right).
\end{align}

Finally, to reproduce the $Y$-dependent part of the solution~(\ref{sol1}), we fix the factor-ordering parameter to
\begin{align}
a=\frac{5}{6},
\end{align}
so that the resulting wave function agrees with the solution obtained by Yeom et al. in this sector.

We now construct wave packets by superposing the mode solutions obtained above. For the particular choice $a=\frac{5}{6}$, the wave function simplifies and can be written as
\begin{align}
\psi(X,Y)\bigg|_{a=\frac{5}{6}}
=\int^{\infty}_{-\infty} f(k)\, e^{-ikX} K_{ik}(2\gamma' e^{Y})\, dk,
\end{align}
where $\gamma'=\frac{r_s\gamma}{2}$ and $f(k)$ is an amplitude function in momentum space. 
Explicitly, one finds
\begin{align}
f(k)=A\!\left(m,\frac{5}{6}\right)
\frac{\sqrt{\pi}\,2^{2+ik}\gamma^{ik}\Gamma\!\left(-\frac{ik}{2}\right)
\sin\!\left(\pi\left(1+\frac{ik}{2}\right)\right)}
{r_s^{\frac{5}{2}-3a}e^{i\frac{\pi}{2}\left(1+\frac{ik}{2}\right)}
\Gamma\!\left(\frac{1}{2}-ik\right)}.
\end{align}
For this solution, as reviewed in the previous section, choosing the amplitude $f(k)$ as in Eq.~(\ref{amplitude1}) leads to the behavior illustrated in Fig.~\ref{fig1}, which realizes Yeom's annihilation-to-nothing scenario.

Owing to the freedom in factor ordering, the Ashtekar–Barbero formulation admits a broader class of quantum solutions than the conventional Wheeler–DeWitt quantization of the corresponding minisuperspace model. More generally, for arbitrary $a$, the wave function takes the form
\begin{align}
\psi(X,Y)=\int^{\infty}_{-\infty} 
f(k)\, e^{-ikX}
e^{\left(-\frac{5}{2}+3a\right)Y}
K_{-\frac{5}{2}+3a-ik}\!\left(2\gamma' e^Y\right) dk.
\end{align}

To analyze the near-horizon behavior, we consider the limit $Y\rightarrow -\infty$. 
In this regime, the modified Bessel function of the second kind admits the asymptotic expansion
\begin{align}
K_{-\frac{5}{2}+3a-ik}(2\gamma' e^{Y})
\simeq \frac{1}{2}\Big(
\gamma'^{-\frac{5}{2}+3a-ik}
e^{-(\frac{5}{2}-3a+ik)Y}
\Gamma\!\left(\frac{5}{2}-3a+ik\right)
\nonumber\\
+\,
\gamma'^{\frac{5}{2}-3a+ik}
e^{(\frac{5}{2}-3a+ik)Y}
\Gamma\!\left(-\frac{5}{2}+3a-ik\right)
\Big).
\end{align}
Substituting this expression into the integral representation of $\psi(X,Y)$, we obtain
\begin{align}
\psi(X,Y)\big|_{Y\rightarrow-\infty}
=\int^{\infty}_{-\infty}
\left(
g(k)\, e^{-ik(X+Y)}e^{-(5-6a)Y}
+
h(k)\, e^{-ik(X-Y)}
\right) dk,
\end{align}
where the coefficient functions are defined by
\begin{align}
g(k)&=\frac{1}{2}
f(k)\gamma'^{-\frac{5}{2}+3a-ik}
\Gamma\!\left(\frac{5}{2}-3a+ik\right),\\
h(k)&=\frac{1}{2}
f(k)\gamma'^{\frac{5}{2}-3a+ik}
\Gamma\!\left(-\frac{5}{2}+3a-ik\right).
\end{align}

In order to explore the quantum behavior of the classical spacetime geometry, we impose a boundary condition such that a Gaussian wave packet is localized at, and propagates into, the horizon $X=Y=-\infty$. This requirement fixes the momentum-space profile to be
\begin{align}
f(k)=
\frac{2C\, e^{-\frac{1}{2}\sigma^2 k^2}}
{\gamma'^{\frac{5}{2}-3a+ik}
\Gamma\!\left(-\frac{5}{2}+3a-ik\right)},
\end{align}
where $C$ is a normalization constant and $\sigma$ denotes the standard deviation 
of the Gaussian packet at $X=Y$.

One immediately observes that for $a<\frac{5}{6}$ the prefactor 
$e^{-(5-6a)Y}$ diverges in the near-horizon limit, implying that no regular solution 
can be defined in this parameter region. 
We therefore restrict attention to solutions satisfying $a\geq\frac{5}{6}$.

With this choice of $f(k)$, the final expression for the wave packet becomes
\begin{align}
\psi(X,Y)=\int^{\infty}_{-\infty}
\frac{2C\, e^{-\frac{1}{2}\sigma^2 k^2}}
{\gamma'^{\frac{5}{2}-3a+ik}
\Gamma\!\left(-\frac{5}{2}+3a-ik\right)}
\, e^{-ikX}
e^{\left(-\frac{5}{2}+3a\right)Y}
K_{-\frac{5}{2}+3a-ik}\!\left(2\gamma' e^Y\right)
dk,
\end{align}
which provides a regular quantum state at the horizon for $a\geq\frac{5}{6}$.

\begin{figure}[htbp]
   \centering
   \includegraphics[width=0.7\linewidth]{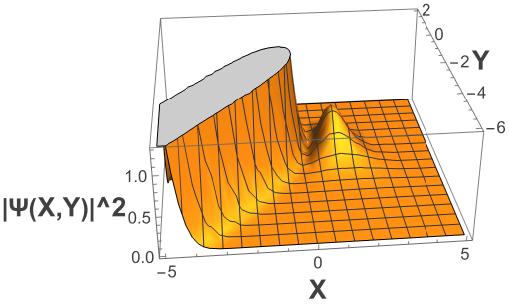}
   \caption{Probability density of the wave function for $C=\sigma=\gamma'=1$, $a=1$. The integral over the wave number $k$ is taken over the interval $[-8,8]$.}
   \label{fig2}
\end{figure}

\begin{figure}[htbp]
   \centering
   \includegraphics[width=0.7\linewidth]{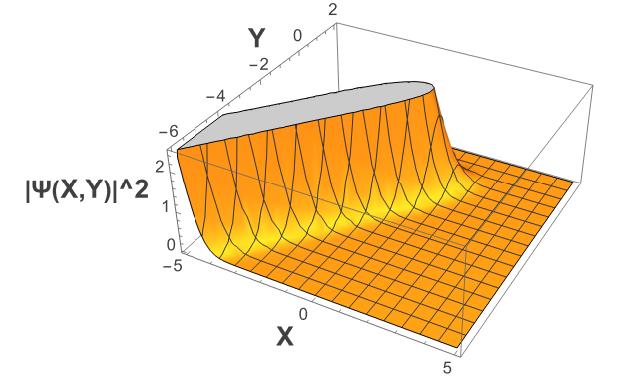}
   \caption{Probability density of the wave function for $C=\sigma=\gamma'=1$, $a=2$. The integral over the wave number $k$ is taken over the interval $[-8,8]$.}
   \label{fig3}
\end{figure}

For these solutions, we select boundary conditions so that a Gaussian wave packet exhibiting classical behavior is localized on the horizon,
and perform a numerical analysis to determine whether the annihilation-to-nothing scenario is realized. The graphical representations of this behavior for the cases $a=1$ and $a=2$ are shown in Figs.~\ref{fig2} and \ref{fig3}. The plots appear to be truncated; this behavior is caused by rapid oscillations of the wave function, which exceed the numerical resolution, and does not indicate any divergence of the function itself. The wave function remains convergent in these regions.

For $a = 1$, the figure shows that the wave packet vanishes at $X = 0$. However, the wave packet is not symmetric with respect to $X = 0$, indicating that it cannot be interpreted as the mutual annihilation of wave packets propagating along opposite arrows of time. This observation may provide a new perspective on the annihilation-to-nothing scenario.

For $a = 2$, on the other hand, the behavior characteristic of the annihilation-to-nothing scenario is not observed in the figure.

\section{Annihilation-to-Nothing Scenario under the Generalized Uncertainty Principle}
\label{GUP}

General relativity is a low-energy effective theory and is therefore not expected to capture physics induced by ultraviolet (UV) dynamics and
additional degrees of freedom responsible for singularity resolution. Accordingly, the annihilation-to-nothing behavior discussed in the previous
sections cannot be reliably assessed without invoking a UV-complete framework.

In this section, we incorporate the generalized uncertainty principle (GUP) as an effective description of UV-originated dynamics and examine its
impact on the annihilation-to-nothing scenario. Since the GUP corrections do not constitute a complete theory of quantum gravity, the present
analysis cannot provide a definitive conclusion regarding the viability of the annihilation-to-nothing scenario. Nevertheless, it is meaningful to
investigate whether the GUP contribution tends to promote or suppress this behavior.

To incorporate minimal uncertainties in the conjugate momenta, we assume
the generalized uncertainty relations
\begin{align}
\label{relation1}
\Delta b \Delta p_b &\ge \frac{G\gamma}{2}
\left(1+\beta_b (\Delta b)^2\right),\\
\label{relation2}
\Delta c \Delta p_c &\ge G\gamma
\left(1+\beta_c (\Delta c)^2\right),
\end{align}
which imply nonvanishing minimal uncertainties in the momenta, of order $G\gamma\sqrt{\beta_b}$ and $2G\gamma\sqrt{\beta_c}$ \cite{Maggiore:1993rv,Kempf:1994su,Scardigli:2003kr,Mignemi:2011wh,Hossenfelder:2012jw,Pramanik:2013zy,Bosso:2020aqm,Wagner:2021bqz,Barca:2021epy,Bosso:2021koi,HerkenhoffGomes:2022bnh,Segreto:2022clx,Bosso:2022rue}. The parameters $\beta_b$ and $\beta_c$ therefore control the strength of
the quantum-gravitational corrections.

Such generalized uncertainty relations can be realized at the operator level by postulating the deformed commutation relations \cite{Pedram:2011aa,Bosso:2020ztk,Bosso:2023sxr,Bosso:2023fnb}
\begin{align}
[b,p_b] &= iG\gamma(1+\beta_b b^2),\\
[c,p_c] &= i2G\gamma(1+\beta_c c^2).
\end{align}
In the limit $\beta_b=\beta_c=0$, the standard canonical algebra is recovered.

To construct an explicit representation of the modified algebra, we introduce auxiliary canonical variables $b_0$ and $c_0$ satisfying
\begin{align}
\label{relation3}
[b_0,p_b] &= iG\gamma,\\
\label{relation4}
[c_0,p_c] &= 2iG\gamma,
\end{align}
and work in the representation where $b_0$ and $c_0$ act multiplicatively
on wave functions. The momentum operators are then realized as
\begin{align}
p_b &= -iG\gamma \frac{\partial}{\partial b_0},\\
p_c &= -2iG\gamma \frac{\partial}{\partial c_0}.
\end{align}

The physical variables $b$ and $c$ can subsequently be expressed as nonlinear functions of $b_0$ and $c_0$ such that the deformed commutation relations are satisfied. Substituting these operators into the Hamiltonian constraint $\hat H\Psi=0$ leads to a deformed Wheeler–DeWitt equation whose structure differs from the standard case through the deformation parameters $\beta_b$ and $\beta_c$.

Furthermore, using the commutation relations (\ref{relation1}), (\ref{relation2}), (\ref{relation3}) and (\ref{relation4}), one can explicitly determine the functional dependence of the physical variables $b$ and $c$ on the auxiliary canonical variables $b_0$ and $c_0$.  A straightforward calculation shows that they are related by
\begin{align}
b &= \frac{1}{\sqrt{\beta_b}} 
\tan\!\left( \sqrt{\beta_b}\, b_0 \right), \\
c &= \frac{1}{\sqrt{\beta_c}} 
\tan\!\left( \sqrt{\beta_c}\, c_0 \right).
\end{align}
This nonlinear transformation ensures that the deformed algebra is satisfied identically. As a consequence of the tangent function, the domains of $b_0$ and $c_0$ are restricted to the finite intervals in Ref~\cite{Bosso:2023sxr}
\begin{align}
-\frac{\pi}{2\sqrt{\beta_b}} < b_0 < \frac{\pi}{2\sqrt{\beta_b}},
\qquad
-\frac{\pi}{2\sqrt{\beta_c}} < c_0 < \frac{\pi}{2\sqrt{\beta_c}}.
\end{align}
Thus, the auxiliary configuration space becomes compact, reflecting the existence of minimal uncertainties in the conjugate momenta.

Having established the explicit form of the variables, we can now rewrite
the Hamiltonian (3.2) in terms of the canonically conjugate pairs
$(b_0,c_0)$ and $(p_b,p_c)$. Acting with the Hamiltonian constraint on a
wave function $\psi_{\mathrm{GUP}}(b_0,c_0)$ in the $(b_0,c_0)$
representation leads to the following differential equation in Ref~\cite{Bosso:2023fnb}:
\begin{equation}
\left[
(b^2+\gamma^2)^{1-a} \frac{\partial}{\partial b_0}
(b^2+\gamma^2)^a
+ 4bc^{1-a} \frac{\partial}{\partial c_0} c^a
\right]
\psi_{\mathrm{GUP}}(b_0,c_0) = 0.
\end{equation}

To solve this equation, we proceed by separation of variables and assume
\begin{align}
\psi_{\mathrm{GUP}}(b_0,c_0) = B(b_0)\, R(c_0).
\end{align}
Substituting this ansatz into the constraint equation and dividing by
$BR$, we obtain two independent ordinary differential equations,
connected by a separation constant $m$.

For the $b_0$-sector, the resulting equation reads
\begin{equation}
\left(
\frac{\tan^2(\sqrt{\beta_b} b_0)}{\beta_b}
+ \gamma^2
\right)
\frac{dB}{db_0}
+ \tan(\sqrt{\beta_b} b_0)\sqrt{\beta_b}
\left(
2a \sec^2(\beta_b b_0) - m
\right)
B(b_0) = 0,
\end{equation}
while for the $c_0$-sector we obtain
\begin{equation}
4 \frac{\tan(\sqrt{\beta_c} c_0)}{\sqrt{\beta_c}}
\frac{dR}{dc_0}
+ \left(
m + 4a \sec^2(\beta_c c_0)
\right)
R(c_0) = 0.
\end{equation}

Both equations can be integrated analytically. The corresponding
solutions are given by
\begin{align}
\label{Bm}
B_m(b_0) &= C_{1m}\,
\sec^2(\beta_b b_0)^{-\frac{m}{2(1-\beta_b\gamma^2)}}
\left(
\frac{\tan^2(\sqrt{\beta_b} b_0)}{\beta_b}
+ \gamma^2
\right)^{-a + \frac{m}{2(1-\beta_b\gamma^2)}}, \\
R_m(c_0) &= C_{2m}\,
\sec^2(\beta_c c_0)^{\frac{m}{8}}
\left(
\frac{\tan(\sqrt{\beta_c} c_0)}{\sqrt{\beta_c}}
\right)^{-a - \frac{m}{4}},
\end{align}
where $C_{1m}$ and $C_{2m}$ are integration constants.

In order to analyze the physical content of these solutions, it is useful
to transform them to the momentum representation. The momentum-space
wave function is defined by the Fourier transform over the compact
domains of $b_0$ and $c_0$,
\begin{align}
\bar{\psi}^{\mathrm{GUP}}_m(p_b,p_c)
&=
\int_{-\frac{\pi}{2\sqrt{\beta_b}}}^{\frac{\pi}{2\sqrt{\beta_b}}}
\int_{-\frac{\pi}{2\sqrt{\beta_c}}}^{\frac{\pi}{2\sqrt{\beta_c}}}
e^{-ip_b b_0} e^{-ip_c c_0}
\psi^{\mathrm{GUP}}_m(b_0,c_0)
\, db_0\, dc_0 .
\end{align}
Owing to the separable form of the wave function, this expression
factorizes as
\begin{align}
\bar{\psi}^{\mathrm{GUP}}_m(p_b,p_c)
= \bar{B}_m(p_b)\, \bar{R}_m(p_c),
\end{align}
where the individual Fourier transforms are given by
\begin{align}
\bar{B}_m(p_b) &=
\int_{-\frac{\pi}{2\sqrt{\beta_b}}}^{\frac{\pi}{2\sqrt{\beta_b}}}
e^{-ip_b b_0} B_m(b_0)\, db_0,  \\
\bar{R}_m(p_c) &=
\int_{-\frac{\pi}{2\sqrt{\beta_c}}}^{\frac{\pi}{2\sqrt{\beta_c}}}
e^{-ip_c c_0} R_m(c_0)\, dc_0.
\end{align}
These momentum-space expressions encode the effects of the generalized uncertainty principle through the compact support and the nonlinear structure inherited from the deformed algebra.

We now evaluate the momentum–space wave functions explicitly. Starting from the integral representation of $\bar{R}_m(p_c)$, the Fourier transform over the compact domain of $c_0$ can be performed analytically. For an arbitrary factor–ordering parameter $a$, the result can be expressed in terms of the regularized hypergeometric function ${}_2\tilde{F}_1(A,B;C;z)={}_2F_1(A,B;C;z)/\Gamma(C)$ as
\begin{align}
\bar{R}(p_c)=&\frac{\pi}{\Gamma\!\left(1+\frac{\lambda}{2}-\frac{m}{8}\right)}
\beta_c^{\frac{1}{8}(4a+m-4)}e^{-\frac{i\pi}{4}(-2+6a+2\lambda+m)}
\left(
e^{i \pi\lambda}
-
e^{\frac{i\pi}{4}(8a+m)}
\right)
\nonumber\\
&\times
\csc\!\left(\frac{\pi}{8}(8a-4\lambda+m)\right)
\Gamma(1+a)
\,{}_2\tilde{F}_1\!\left(
1+a,\,
\frac{1}{8}\!\left(8-4\lambda-m\right);\,
\frac{1}{8}\!\left(8+8a-4\lambda+m\right);\,
-1
\right),
\end{align}
where the dimensionless parameter
\begin{align}
\lambda \equiv \frac{r_s^2 e^{2Y-2X}}{\sqrt{\beta_c}},
\end{align}
has been introduced for compactness.

For particular choices of the separation constant $m$,
which correspond to specific values of the factor–ordering parameter,
the above expression simplifies considerably.
For instance, taking $m=\frac{2}{3}+2ik$ yields
\begin{align}
\bar{R}_{\frac{2}{3}+2ik}=&
\frac{\pi\Gamma\!\left(\frac{11}{6}\right)}
{\Gamma\!\left(
\frac{11}{12}+\frac{\lambda}{2}-\frac{ik}{4}
\right)}
\beta_c^{\frac{1}{8}(4a+m-4)}
e^{-\frac{i\pi}{4}\left(\frac{11}{3}+2\lambda+2ik\right)}
\left(
e^{i \pi\lambda}
-
e^{\frac{i\pi}{4}\left(\frac{22}{3}+2ik\right)}
\right)
\nonumber\\
&\times
\csc\!\left(
\frac{\pi}{8}\left(\frac{22}{3}-4\lambda+2ik\right)
\right)
\,{}_2\tilde{F}_1\!\left(
\frac{11}{6},\,
\frac{1}{8}\!\left(\frac{22}{3}-4\lambda-2ik\right);\,
\frac{1}{8}\!\left(\frac{46}{3}-4\lambda+2ik\right);\,
-1
\right).
\end{align}

Similarly, for $m=2ik$ one finds
\begin{align}
\bar{R}_{2ik}=&
\frac{\pi}{\Gamma\!\left(1+\frac{\lambda}{2}-\frac{ik}{4}\right)}
\beta_c^{\frac{1}{8}(4a+m-4)}
e^{-\frac{i\pi}{4}\left(4+2\lambda+2ik\right)}
\left(
e^{i \pi\lambda}
-
e^{\frac{i\pi}{4}\left(8+2ik\right)}
\right)
\nonumber\\
&\times
\csc\!\left(
\frac{\pi}{8}\left(8-4\lambda+2ik\right)
\right)
\,{}_2\tilde{F}_1\!\left(
2,\,
\frac{1}{8}\left(8-4\lambda-2ik\right);\,
\frac{1}{8}\left(16-4\lambda+2ik\right);\,
-1
\right),
\end{align}
while for $m=-4+2ik$ the result becomes
\begin{align}
\bar{R}_{-4+2ik}=&
\frac{2\pi}
{\Gamma\!\left(
1+\frac{\lambda}{2}+\frac{1}{8}(4-2ik)
\right)}
\beta_c^{\frac{1}{8}(4a+m-4)}
e^{-\frac{i\pi}{4}\left(6+2\lambda+2ik\right)}
\left(
e^{i \pi\lambda}
-
e^{\frac{i\pi}{4}\left(12+2ik\right)}
\right)
\nonumber\\
&\times
\csc\!\left(
\frac{\pi}{8}\left(12-4\lambda+2ik\right)
\right)
\,{}_2\tilde{F}_1\!\left(
3,\,
\frac{1}{8}\left(12-4\lambda-2ik\right);\,
\frac{1}{8}\left(20-4\lambda+2ik\right);\,
-1
\right).
\end{align}

An analogous procedure can be followed in the $b$–sector. Performing the Fourier transform of $B_m(b_0)$, the resulting expressions can be written in terms of the regularized generalized hypergeometric function ${}_3\tilde{F}_2$. For arbitrary $a$ and $m$, the Fourier transform of
$B_m(b_0)$ defined in Eq.~(\ref{Bm}) can be evaluated as
\begin{align}
\bar{B}_m(p_b)
&=
\int_{-\frac{\pi}{2\sqrt{\beta_b}}}^{\frac{\pi}{2\sqrt{\beta_b}}}
e^{-i b_0 p_b}
\left[\sec^{2}\!\left(\beta_b b_0\right)\right]^{-\frac{m}{2(1-\beta_b\gamma^2)}}
\left(
\frac{\tan^{2}\!\left(\sqrt{\beta_b} b_0\right)}{\beta_b}
+\gamma^2
\right)^{-a+\frac{m}{2(1-\beta_b\gamma^2)}}
\, db_0
\nonumber\\
&=
\frac{\pi}{2}\,
\beta_b^{-\frac{1}{2}+a+\frac{m}{2\beta_b\gamma^2-2}}\,
{}_3\tilde{F}_2\!\left(
1,\,\frac{3}{2},\,a+\frac{m}{2\beta_b\gamma^2-2}\;;\;
\frac{1}{2}\!\left(3-\frac{p_b}{\sqrt{\beta_b}}\right),\,
\frac{1}{2}\!\left(3+\frac{p_b}{\sqrt{\beta_b}}\right)\;;\;
1-\beta_b\gamma^2
\right).
\end{align}
These expressions explicitly encode the effects of the generalized
uncertainty principle through their dependence on $\beta_b$ and
$\beta_c$, and reduce to the standard results in the limit
$\beta_b,\beta_c \to 0$.
For example, for $m=\frac{2}{3}+2ik$ one obtains
\begin{align}
\bar{B}_{\frac{2}{3}+2ik}=
\frac{\pi}{2}\,
\beta_b^{\frac{1}{3}+\frac{\frac{2}{3}+2ik}{2\beta_b\gamma^2-2}}\,
{}_3\tilde{F}_2\!\left(
1,\,\frac{3}{2},\,\frac{5}{6}+\frac{\frac{2}{3}+2ik}{2\beta_b\gamma^2-2}\;;\;
\frac{1}{2}\!\left(3-\frac{p_b}{\sqrt{\beta_b}}\right),\,
\frac{1}{2}\!\left(3+\frac{p_b}{\sqrt{\beta_b}}\right)\;;\;
1-\beta_b\gamma^2
\right),
\end{align}
whereas for $m=2ik$ we find
\begin{align}
\bar{B}_{2ik}=
\frac{\pi}{2}\,
\beta_b^{\frac{1}{2}+\frac{2ik}{2\beta_b\gamma^2-2}}\,
{}_3\tilde{F}_2\!\left(
1,\,\frac{3}{2},\,1+\frac{2ik}{2\beta_b\gamma^2-2}\;;\;
\frac{1}{2}\!\left(3-\frac{p_b}{\sqrt{\beta_b}}\right),\,
\frac{1}{2}\!\left(3+\frac{p_b}{\sqrt{\beta_b}}\right)\;;\;
1-\beta_b\gamma^2
\right),
\end{align}
and for $m=-4+2ik$,
\begin{align}
\bar{B}_{-4+2ik}=
\frac{\pi}{2}\,
\beta_b^{\frac{3}{2}+\frac{-4+2ik}{2\beta_b\gamma^2-2}}\,
{}_3\tilde{F}_2\!\left(
1,\,\frac{3}{2},\,2+\frac{-4+2ik}{2\beta_b\gamma^2-2}\;;\;
\frac{1}{2}\!\left(3-\frac{p_b}{\sqrt{\beta_b}}\right),\,
\frac{1}{2}\!\left(3+\frac{p_b}{\sqrt{\beta_b}}\right)\;;\;
1-\beta_b\gamma^2
\right).
\end{align}

The solutions incorporating the GUP corrections reproduce the results of the previous section in the limit $\beta_b, \beta_c \to 0$. Therefore, in the infrared (IR) regime where the GUP contributions are expected to be negligible, in particular near the horizon, the solutions should reduce to those obtained previously. Accordingly, We discuss the annihilation-to-nothing scenario using the following solution constructed with the amplitude function $F(k,a)$, which is chosen so that the solution behaves as a Gaussian wave packet on the horizon, as obtained in the previous section.
\begin{align}
\psi^{\mathrm{GUP}}(X,Y,a)=\int^{\infty}_{-\infty}F(k,a)\bar{B}_{4-4a+2ik}(p_b)\bar{R}_{4-4a+2ik}(p_c)dk,
\end{align}
where
\begin{align}
F(k,a)=\frac{2C\, e^{-\frac{1}{2}\sigma^2 k^2}}
{\gamma'^{\frac{5}{2}-3a+ik}
\Gamma\!\left(-\frac{5}{2}+3a-ik\right)}\frac{r_s^{\frac{5}{2}-3a}e^{i\frac{\pi}{2}\left(1+\frac{ik}{2}\right)}
\Gamma\!\left(\frac{1}{2}-ik\right)}{\sqrt{\pi}\,2^{2+ik}\gamma^{ik}\Gamma\!\left(-\frac{ik}{2}\right)
\sin\!\left(\pi\left(1+\frac{ik}{2}\right)\right)}.
\end{align}

\begin{figure}[htbp]
   \centering
   \includegraphics[width=0.7\linewidth]{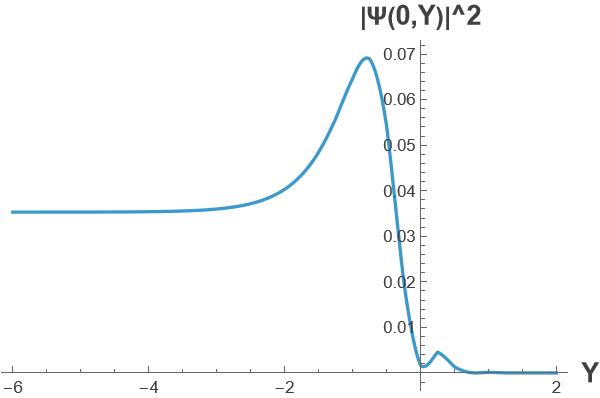}
   \caption{Probability density of the wave function for $a=\frac{5}{6},\gamma=\frac{r_s}{2}=2\beta_b=\beta_c=1$.}
   \label{fig4}
\end{figure}

\begin{figure}[htbp]
   \centering
   \includegraphics[width=0.7\linewidth]{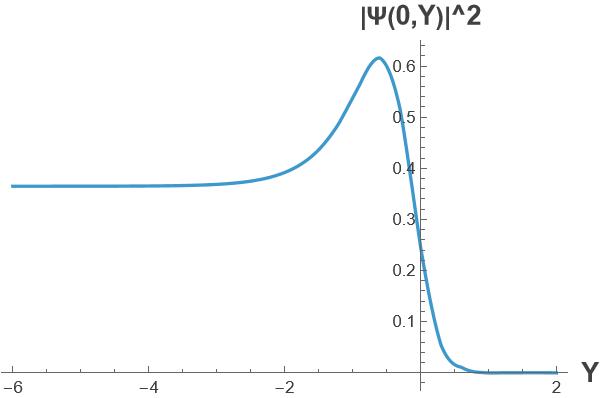}
   \caption{Probability density of the wave function for $a=1,\gamma=\frac{r_s}{2}=2\beta_b=\beta_c=1$.}
   \label{fig5}
\end{figure}

As in the previous section, we repeat the same analysis in the presence of the GUP corrections. Imposing boundary conditions such that a Gaussian wave packet exhibiting classical behavior is localized on the horizon, we numerically investigate whether the annihilation-to-nothing scenario is realized once the UV-modified dynamics is taken into account. Since the functional form becomes considerably more involved in the present case, we do not present two-dimensional plots. Instead, we show one-dimensional profiles evaluated at $X=0$. The corresponding results for $a=\frac{5}{6}$, $a=1$ and $a=2$ are shown in Figs.~\ref{fig3}, \ref{fig4} and \ref{fig5}. In all figures, the integral over the wave number $k$ is taken over the interval $[-8,8]$.

\begin{figure}[htbp]
   \centering
   \includegraphics[width=0.7\linewidth]{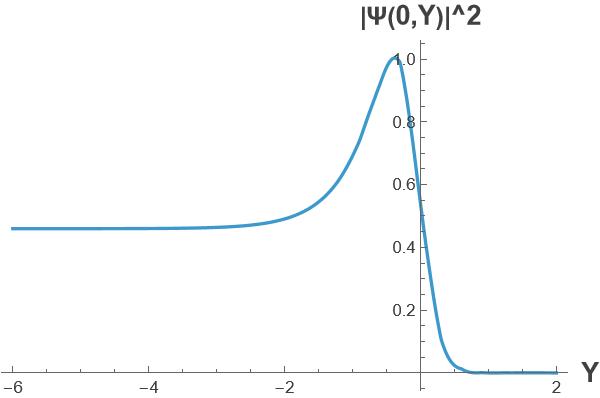}
   \caption{Probability density of the wave function for $a=2,\gamma=\frac{r_s}{2}=2\beta_b=\beta_c=1$. }
   \label{fig6}
\end{figure}

From Figs.~\ref{fig3}, \ref{fig4}, and \ref{fig5}, we find that the mutual annihilation picture of wave packets characteristic of the annihilation-to-nothing scenario is not observed for any value of $a$. In the coordinate system $(X,Y)$, the classical region is given by $Y \geq \ln 2$. In Figs.~\ref{fig3}, \ref{fig4}, and \ref{fig5}, the probability density vanishes for $Y > \ln 2\approx0.693$, indicating that the quantum creation of spacetime is suppressed in this region.

\FloatBarrier
\section{Summary and Discussion}
\label{summary}

In this work, we have investigated the quantum dynamics of the Schwarzschild black hole interior in the Ashtekar–Barbero formulation, with particular emphasis on the fate of the classical singularity and the viability of the annihilation-to-nothing scenario. Exploiting the isometry between the Schwarzschild interior and the Kantowski–Sachs cosmology, we constructed the reduced Hamiltonian in connection–triad variables and performed Wheeler–DeWitt quantization in minisuperspace.

In the standard Schrödinger representation, we derived exact separable solutions of the Hamiltonian constraint and analyzed the role of factor ordering. We showed that the annihilation-to-nothing behavior can be reproduced only for a specific choice of ordering parameter. For more general orderings, the wave packet either loses the time-symmetric structure required for mutual annihilation or fails to exhibit the characteristic cancellation of branches. This indicates that the scenario is not generic within the Ashtekar–Barbero framework and depends sensitively on quantization ambiguities.

Furthermore, following existing formulations of the generalized uncertainty principle, we incorporated ultraviolet-motivated corrections. In this framework, the canonical algebra is deformed, leading to minimal uncertainties in the conjugate momenta. By adopting the representation in terms of auxiliary canonical variables introduced in previous works, we constructed an explicit realization of the deformed algebra and derived the corresponding modified Wheeler–DeWitt equation. The resulting solutions are expressed in terms of generalized hypergeometric functions and encode minimal-length effects through the compact domains of the configuration variables and the nonlinear operator structure.

Imposing Gaussian boundary conditions at the horizon, we numerically analyzed the resulting wave packets. Our results show that once the GUP corrections are included, the annihilation-to-nothing behavior is suppressed for all factor orderings considered. The mutual annihilation of oppositely oriented classical branches is not realized in the ultraviolet-modified dynamics. In this sense, the annihilation-to-nothing scenario does not appear to be robust under minimal-length deformations of the canonical structure.

At the same time, the GUP modification alters the ultraviolet behavior of the wave function and affects the structure of the quantum geometry near the classical singularity. Although the present analysis does not constitute a complete theory of quantum gravity, it demonstrates that minimal-length effects can qualitatively change the interior quantum dynamics. This suggests that conclusions drawn from the undeformed Wheeler–DeWitt framework should be treated with caution when ultraviolet corrections are taken into account.

Several issues deserve further investigation. First, it would be important to analyze expectation values of curvature operators more systematically in the deformed framework in order to clarify whether singularity resolution occurs in a stronger sense. Second, the relation between the present approach and loop quantum gravity–inspired polymer quantization \cite{Ashtekar:2002sn,Corichi:2007tf,Morales-Tecotl:2015vto,Morales-Tecotl:2016ijb} remains to be explored. Finally, extending the analysis beyond minisuperspace, or incorporating matter fields, may provide additional insight into the robustness of boundary-condition–based singularity resolution mechanisms.

We hope that the present results contribute to a more precise understanding of the role of quantization ambiguities and minimal-length effects in black hole interior dynamics and in the broader problem of spacetime singularities in quantum gravity.

\bibliography{references}

\end{document}